\documentclass[conference]{IEEEtran}
\newif\ifarxivversion
\arxivversiontrue

\IEEEoverridecommandlockouts
\usepackage{amsmath,amssymb,amsfonts}
\usepackage{algorithmic}
\usepackage{graphicx}
\usepackage{textcomp}
\usepackage[table,xcdraw]{xcolor}
\usepackage{multirow}
\usepackage{booktabs}
\usepackage{balance}
\usepackage[hidelinks]{hyperref}

\usepackage[backend=biber, abbreviate=true, dateabbrev=true, alldates=year, isbn=false, doi=false, urldate=comp, url=false, maxbibnames=5, backref=false, style=numeric-comp, language=american, giveninits=true, sortcites=true, sorting=none]{biblatex}
\AtEveryBibitem{\clearlist{location}} %
\AtEveryBibitem{\clearfield{series}} %
\AtEveryBibitem{\clearlist{language}} %

\usepackage[table]{xcolor}
\usepackage{pgf} %

\newcommand{\defnote}[2]{%
  \refstepcounter{footnote}%
  \label{#1}%
  \footnotetext{#2}%
}

\newcommand{\SoBigDataIPAck}{European Union – Horizon EUROPE Work Program under the scheme "HORIZON-INFRA-2025-01-DEV-02 - Developing, consolidating and optimising the European research infrastructures landscape, maintaining global leadership (INFRADEV)", Grant Agreement n.101292277, “SoBigData IP: SoBigData Implementation Phase"\xspace}

\newcommand{\ArxivConferenceNotice}{%
  \begingroup\footnotesize%
  \setlength{\parindent}{0pt}%
  \rule{\columnwidth}{0.4pt}\\[0.45em]%
  \begin{minipage}[c]{0.2\columnwidth}
    \includegraphics[width=\columnwidth]{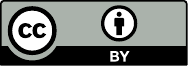}%
  \end{minipage}\hspace{4pt}%
  \begin{minipage}[c]{0.65\columnwidth}\raggedright%
    This work is licensed under a Creative Commons Attribution 4.0 International (CC BY 4.0) license.
  \end{minipage}\\[.4em]
  \begin{minipage}{\columnwidth}%
    Copyright \copyright\ 2026 held by the owner/author(s).\\[.2em]
    \emph{Accepted for publication in the Research Papers Track of the 42nd IEEE International Conference on Software Maintenance and Evolution (ICSME~2026),\newline 14--18~September 2026, Benevento, Italy.}%
  \end{minipage}%
  \endgroup
}

\usepackage{tikz,pifont}
\usepackage{amsfonts,amsmath}
\usepackage{hyperref}
\usepackage{xspace}
\usepackage{multirow}
\usepackage{makecell}
\usepackage{enumitem}
\usepackage{booktabs}
\usepackage[many]{tcolorbox}

\newlist{rqs}{enumerate}{1}
\setlist[rqs,1]{label=\textbf{RQ\arabic*.},ref=\textbf{RQ\arabic*}}

\makeatletter
\let\MYcaption\@makecaption
\makeatother
\usepackage[font=footnotesize]{subcaption}
\makeatletter
\let\@makecaption\MYcaption
\makeatother

\newtcolorbox{rqanswer}{
    colback=gray!25,
    colframe=black, 
    enhanced,
    boxrule=0mm,
    frame hidden,
    left=1mm, 
    top=1mm, bottom=1mm, right=1mm,
    borderline west={0.8mm}{0mm}{gray!80!black}, 
    sharp corners, 
} 
\begin{document}

\thispagestyle{plain}
\pagestyle{plain}

\title{Repair on a Budget: An Empirical Study on LLM Quantization for Automated Program Repair}

\title{Smaller Models, Unexpected Costs: Trade-offs in LLM Quantization for Automated Program Repair}

\author{\IEEEauthorblockN{Fernando Vallecillos-Ruiz\IEEEauthorrefmark{1}, Giordano d'Aloisio\IEEEauthorrefmark{2}, Max Hort\IEEEauthorrefmark{1}, Luca Traini\IEEEauthorrefmark{2}, Antinisca Di Marco\IEEEauthorrefmark{2}, Leon Moonen\IEEEauthorrefmark{1}}
\IEEEauthorblockA{\IEEEauthorrefmark{1}Simula Research Laboratory, Oslo, Norway,
\{fernando,maxh\}@simula.no, leon.moonen@computer.org\\
\IEEEauthorrefmark{2}University of L'Aquila, L'Aquila, Italy,
\{giordano.daloisio,luca.traini,antinisca.dimarco\}@univaq.it
}
}

\maketitle

\begin{abstract}

Large Language Models (LLMs) are powerful tools and have been increasingly adopted for complex software engineering tasks.
As the number of parameters increases, results can often be improved, but this also imposes substantial memory requirements.
While quantization effectively reduces the memory footprint,
its overall impact is often summarized only by benchmark scores, which mask changes in model behavior and non-functional overheads.
In this work, we conduct an empirical evaluation of LLM quantization using Automated Program Repair (APR), a complex task in software engineering.
We analyze 13 quantization configurations spanning different bit-widths, methods, and target components (weights and KV-cache) across six representative LLMs, evaluated on two APR benchmarks (HumanEval-Java and Defects4J).

Our findings reveal that base and quantized models can provide different sets of repaired problems with little overlap, while retaining a comparable number of repaired problems.
Although quantization successfully reduces memory footprints by up to 85\%, it increases both inference time and energy consumption, which we attribute to suboptimal hardware utilization.
Our Pareto trade-off analysis shows that 48\% of the configurations evaluated are strictly dominated by alternatives.
Rather than identifying a superior quantization method, our findings highlight that the trade-offs between 
effectiveness,
memory footprint, and energy efficiency are sensitive to the underlying model architecture and the complexity of the task.

\end{abstract}


\section{Introduction}
\ifarxivversion
\begin{figure}[b]
  \ArxivConferenceNotice
\end{figure}
\fi
Large language models (LLMs) are powerful tools and have been applied to a variety of software engineering tasks~\cite{hou2024large,fan2023large}. 
However, the increasing model size
imposes substantial hardware constraints.
Deploying models locally or integrating them into existing pipelines both requires a substantial amount of memory and incurs significant operational costs.
As such, techniques have been proposed to improve their efficiency along dimensions such as inference time, energy consumption, and memory size~\cite{zhou2024survey}.
Among these, quantization~\cite{gholami2022survey} is popular for reducing memory requirements,
a bottleneck in most environments. 
In the case of LLMs, quantization lowers the precision of weights, for instance from 32-bit floating point to 8-bit, and thereby is able to reduce the 
memory footprint.

Generally, code generation has been a popular application for LLM quantization~\cite{melin2024:precision,afrin2025:quantization,nyamsuren2025:evaluating,gong2024:llmc,shi2025:systematic,wei2023:greener,giagnorio2025:quantizing}.
However, the use of quantization methods for other code-related tasks, such as Automated Program Repair (APR), has received less focus, and the cost and benefits of quantization for APR are yet to be studied in further detail.
Specifically, existing studies are limited to one quantization method, a single benchmark, or focus on one type of efficiency metric (i.e., memory~\cite{kusama2025:how}, energy~\cite{alizadeh2025:language}).

We aim to fill this gap by providing a comprehensive evaluation of 13 quantization configurations derived from five different quantization methods, for six LLMs (ranging from 6.7B to 70B parameters), for two APR benchmarks.
We use default recommended settings to reflect realistic out-of-the-box usage by most practitioners.
In addition to evaluating the model effectiveness (a.k.a. solution plausibility and solved-set consistency between base and quantized models), we include three efficiency dimensions (i.e., inference time, energy consumption, memory footprint), and study trade-offs between effectiveness and efficiency in quantized and base models. 

Our results show that quantization can yield substantial memory savings, but these do not translate to lower inference time or energy consumption.
Rather than trying to identify a universally optimal configuration, we analyze the trade-offs between repair capability, solved-set consistency, and efficiency depending on factors such as model family, quantization target, and bit-width.

To summarize, we make the following contributions:
\begin{itemize}
    \item We conduct an extensive empirical study of 13 quantization configurations 
    spanning weight-only and KV-cache settings from 2 to 8 bits
    across
    six LLMs ranging from 6.7B to 70B parameters on two APR benchmarks. 
    \item We assess the bug repair ability of quantized LLMs based on test pass rate 
    and on Jaccard Consistency Rate (JCR), a new metric we propose to assess the solved-set consistency between base and quantized models, helping us to reveal that similar plausibility counts can mask substantial shifts in which bugs a model is able to repair.
    \item We quantify the efficiency of quantization methods with regard to three non-functional metrics (i.e., inference time, energy consumption, memory footprint), reporting non-functional effects using bootstrapped confidence intervals for relative changes.
    \item We analyze effectiveness--efficiency trade-offs through Pareto dominance and show that, in our study, 48\% of the evaluated quantization configurations are strictly dominated by alternative settings.
    \item We share a replication package containing all scripts for carrying out experiments, as well as the generated patches by the LLMs with all efficiency metrics.\footnote{~\url{https://doi.org/10.5281/zenodo.20847818}\label{footnote:repl-package}}
\end{itemize} 
\section{Related Work}\label{sec:related}
\subsection{Quantization for LLMs}
Quantization is a popular compression strategy which reduces the precision of numbers, such as model weights~\cite{gholami2022survey}, with the goal of reducing memory footprint and inference costs. 
For instance, full precision floating point numbers (e.g., 32-bit) in LLMs can be reduced to a lower precision (e.g., 8-bit integers) while aiming at preserving the same level of performance.
The two main approaches for quantization are Quantization-Aware Training (QAT) and Post-Training Quantization (PTQ)~\cite{wang2024:art}.
Here, PTQ describes approaches that are applied once a model has been trained and converts floating point numbers 
to lower precision without retraining.
QAT, on the other hand, performs quantization during the training 
of an LLM, enabling models to manage quantization noise. 
While PTQ is more susceptible to noise than QAT, it benefits from its ease of use and lower computational costs~\cite{daloisio2024:compression}.

Beyond this classification on when the quantization is applied, quantization methods can also be categorized by their target component~\cite{shi2025:systematic,liu2025:quantization,lang2024:comprehensive}:
\begin{itemize}
    \item Weight-only: only weights are quantized. 
    Weight-only PTQ is widely adopted by practitioners due to its easy applicability and memory savings.
    \item Activation (or weight-activation): weights and activations are quantized (e.g., W8A8 or W4A8).
    This approach yields larger memory saving and compute speedups but requires careful handling of activation outliers~\cite{xiao2023:smoothquant}.
    \item KV Cache: KV cache is quantized. During the decoding stage, LLMs store KV Cache to avoid recomputing attention. 
    KV cache grows with context and can dominate memory usage for long contexts.
\end{itemize}

\subsection{Quantization in SE}
An early study on quantization for software engineering tasks with LLMs 
was conducted by Wei et al.~\cite{wei2023:greener} in 2023.
They investigated the impact of quantization on four LLMs (PLBART, Code-T5, InCoder, CodeGen) for Python code generation. 
In particular, the effect of quantization is measured along three dimensions: resource usage, accuracy, and robustness.
For this purpose, static and dynamic post-training quantization are used to quantize weights and activations from fp32 down to 4-bit integers.
Results showed that quantized models benefit from lower resource usage (e.g., inference time, memory and carbon footprint), with minimal accuracy loss.

Giagnorio et al.~\cite{giagnorio2025:quantizing} carried out an extension of the study by Wei et al.~\cite{wei2023:greener}, considering the code generation task for Python and Java. 
This extension added recent models, with sizes up to 34B parameters, and novel quantization methods allowing for 2-bit quantization.
Specifically, they used the Additive Quantization with Learned Multi-Codebooks~\cite{egiazarian2024:extreme} PTQ quantization method and with varying quantization degrees (i.e., 8, 4, 3, 2 bits).
Moreover, they found that effectiveness degradation due to quantization can be mitigated by incorporating calibration datasets that include code. 

While code generation is the most popular application of LLM quantization in the software engineering domain~\cite{melin2024:precision,afrin2025:quantization,nyamsuren2025:evaluating,gong2024:llmc,shi2025:systematic}, quantization has also been applied to other tasks.
These include vulnerability detection, code summarization, code search~\cite{daloisio2024:compression}, code
summarization~\cite{wei2023:greener}, and program repair~\cite{kusama2025:how,alizadeh2025:language}.

For APR, Kusama et al.~\cite{kusama2025:how} investigated quantization with GPTQ, reducing float32 to int8, and evaluating the performance of Phi and Qwen2.5-Coder models of different sizes.
Alizadeh et al.~\cite{alizadeh2025:language} followed an energy perspective to investigate the performance and energy consumption of LLMs and quantized variants on four tasks, one of which was program repair. 
Their results showed that larger models tend to consume more energy than their quantized counterparts, while no clear trend for accuracy is observed. %

While both works provide insights on quantization for APR, they considered a single benchmark and quantization method, as well as focussing on a single efficiency dimension (e.g., memory~\cite{kusama2025:how}, energy~\cite{alizadeh2025:language}).
We extend the investigation of quantization for APR by considering two challenging benchmarks, five quantization methods with different reduction sizes, as well as three efficiency dimensions and trade-offs.

\subsection{Evaluating Quantization}
LLMs achieved high effectiveness in several tasks, but with the reduction of weight precision, quantization risks the loss of performance. 
Therefore, one of the main goals of quantization methods is to maintain the performance level of the base model~\cite{wang2024fp4,hershcovitch2024lossless}.
While it is important that the impact of quantization on performance metrics (e.g., accuracy) is minimal, it should not be looked at in a vacuum.
In particular, metrics such as accuracy can hide differences in the decision-making process between the base and quantized models. 
For instance, two models achieving an accuracy of 50\% for a binary prediction task might disagree in all cases. 
Therefore, it is necessary to investigate how often models make identical decisions~\cite{kamal2025:downsized}, in addition to determining whether they are correct or not.
This ensures that faithfulness to the initial model is maintained, which is crucial to ensure that quantization does not lead to issues related to safety and robustness~\cite{lin2019:defensive,fang2025:smaller,wei2023:greener}.

To the best of our knowledge, Okuno et al.~\cite{okuno:lossless} are the first to address this issue, and describe it with the term ``lossless AI''.\footnote{~We note that this term is commonly used differently, to describe scenarios without accuracy losses.}
Notably, they required that compressed models make predictions consistent with the respective larger, reference model.
For the task of image classification, they specified three inconsistencies between predictions made by a reference model and its compressed variant:
(i) degradation, where the reference is correct but the compressed model is not; (ii) improvement, where only the compressed model is correct; and (iii) changes between incorrect classes, where both models make different wrong predictions.
    
In addition to accuracy, Hu et al.~\cite{hu2023:understanding} reported the number of disagreements to quantify behaviour changes after quantization when dealing with image and text classification tasks under different network architectures.
Besides consistency in classifications, Li et al.~\cite{li2025:efficiency} determined the similarity of generated images by quantized models. 

When it comes to measuring the efficiency impact of quantization methods for SE tasks, several dimensions have been considered. 
Naturally, the size of the models after quantization is being reported~\cite{kusama2025:how,giagnorio2025:quantizing,daloisio2024:compression,wei2023:greener}, while the work by Wei et al.~\cite{wei2023:greener} also reported the memory required for the inference process. 
Other popular aspects include inference time~\cite{nyamsuren2025:evaluating,daloisio2024:compression,wei2023:greener}
or the sustainability of quantization (e.g., energy efficiency~\cite{alizadeh2025:language,shi2025:systematic}, $CO_2$ emissions~\cite{wei2023:greener}).
Lastly, the trade-off between task effectiveness and efficiency 
with different quantization methods is explicitly analyzed in several works~\cite{nyamsuren2025:evaluating,giagnorio2025:quantizing,daloisio2024:compression,alizadeh2025:language,shi2025:systematic}.

\section{Methodology}\label{sec:methodology}

Building on the gaps identified above, we now describe the design of our empirical study.
The \emph{goal} of our study is to analyze the impact of quantization methods on the \textit{effectiveness} 
(plausibility and consistency) and \textit{efficiency}
(inference time, energy consumption, and memory footprint) of LLMs for APR. Specifically, our research is driven by the following research questions (RQ):

\begin{itemize}[leftmargin=*]
    \item \textbf{RQ1. Effectiveness:} How does quantization affect the automated program repair capability of LLMs?
    \begin{itemize}
        \item How does quantization affect the number of problems with at least one plausible patch?
        \item To what extent does quantization preserve the set of problems that a model is able to repair?
    \end{itemize}
    \item \textbf{RQ2. Efficiency:} How does quantization affect the \textit{i)} inference time, \textit{ii)} energy consumption, and \textit{iii)} memory footprint of LLMs for APR? 
    \item \textbf{RQ3. Effectiveness-Efficiency Trade-offs:} What trade-offs emerge between APR effectiveness and efficiency when adopting quantized LLM variants, and how do different quantization configurations balance these objectives?
\end{itemize}

\subsection{Quantization Methods}
We focus on post-training quantization (PTQ), which compresses existing pre-trained models.
Some of the methods employ a calibration phase 
with a small number of samples
~\cite{wang2024:art,shi2025:systematic,daloisio2024:compression}.
We divide the methods according to the target component they quantize: model weights, or KV cache.
For each component, we employ different methods with a variety of precision levels.
Since each method and implementation exposes different precision levels, we do not evaluate the same for every method.
The concrete choices are reported below.

\subsubsection{Model weight quantization}
In this set of methods, the model's weights are replaced by low-bit representations.
We include five quantization methods to cover different algorithmic and runtime implementations:

\textbf{Additive Quantization of Language Models (AQLM)}~\cite{egiazarian2024:extreme} is a data-aware PTQ method that generalizes additive quantization to LLMs by encoding each weight group as the sum of multiple learned codebooks rather than a fixed one.
AQLM optimizes its codebooks against behaviour on calibration data which makes it suitable for extreme compression (2--3 bits per parameter).
However, it is more computationally expensive than on-the-fly approaches.

\textbf{Activation-aware Weight Quantization (AWQ)} ~\cite{lin2025:awq} is a lighter-weight form of data-aware quantization.
It uses calibration data to identify channels that are most sensitive to quantization error and searches per-channel rescaling factors that protect salient weights before quantizing.
Since it does not rely on reconstruction nor backpropagation, AWQ preserves generalizability without overfitting to the calibration data.

\textbf{BitsAndBytes (BNB)}~\cite{dettmers2022:llmint8} is a practical quantization method used in the Hugging Face ecosystem replacing standard linear layers with lower bit modules.
It is calibration-free, offering 8- and 4-bit quantization, from which we select 4-bit precision.
We include it in our study because it is widely used and easy to integrate into most pipelines.

\textbf{Half-Quadratic Quantization (HQQ)}~\cite{badri2023:halfquadratic} applies half-quadratic optimization to fit a low-bit representation of the model's weights.
It quantizes weights directly and does not require a separate calibration dataset.
We select HQQ for its flexibility, as it supports several bitwidths (8, 4, 3, 2, 1-bit), from which we select the first three.
The latter two bitwidths, while possible, resulted in unusable outputs by all models.

\textbf{Quanto}~\cite{huggingface2024:quanto} is a quantization method integrated in the Hugging Face environment and supporting low-bit integer weight formats for inference.
It is also a calibration-free method offering 8, 4, and 2-bit quantization from which we discard the last one on the same basis as with HQQ.

\subsubsection{KV cache quantization}
We also evaluate \textbf{HQQ} and \textbf{Quanto} applied to KV cache only.
This setting keeps model weights unchanged and reduces precision only for cached attention keys and values at the decoding stage.
It targets inference-time memory growth with sequence length rather than static model footprint.
Implementation-wise, KV cache quantization is enabled through the Transformers quantized-cache mechanism with method-specific cache configurations recommended in the documentation.
In our setup, Quanto KV-cache is evaluated at 2 and 4 bits, while HQQ KV-cache is evaluated at 2, 4, and 8 bits.
For each technique, we follow the recommended parameters established in their documentation.

\medskip

For notational convenience, we use subscripts to denote the bit-width precision employed, while the prediction target is reported in parentheses. For instance, hqq$_8$ (M) denotes a configuration where half-quadratic quantization is applied to the model weights at 8-bit precision, while quanto$_4$ (KV) denotes a configuration where the precision of the KV cache is reduced to 4 bits using Quanto.

\subsection{Models}
For our empirical evaluation, we apply PTQ quantization to six pre-trained LLMs (Table~\ref{tab:overview-models}).
In particular, we considered three different model families, and we select models of two different sizes (i.e., one in the range of 6-8 billion parameters, and one in the range from 30-70 billion parameters). 
The three considered model families are DeepSeek-Coder~\cite{guo2024:deepseekcoder}, Llama-3~\cite{dubey2024:llama} and Mistral~\cite{jiang2023:mistral,jiang2024:mixtral}, which have shown their coding ability in previous quantization works~\cite{afrin2025:quantization,melin2024:precision}.

For quantization methods that require calibration, we employ publicly available calibrated LLM versions. These versions are indicated in our replication package.

\begin{table}[t]
\vspace*{.5ex}
\centering
\caption{Overview of language models used.}
\label{tab:overview-models}

\begin{tabular}{@{}llrc}
\toprule
\textbf{Family} & \textbf{Model Name}                    & \textbf{Size} \\ \midrule \midrule
Llama       & Llama-3-8B-Instruct\footref{foot:model-cl-s}                 & 8B                            \\
                & Llama-3-70B-Instruct\footref{foot:model-cl-l}                    & 70B                               \\ \midrule
DeepSeek        & deepseek-coder-6.7b-instruct\footref{foot:model-ds-s}              & 6.7B                            \\
                & deepseek-coder-33b-instruct\footref{foot:model-ds-l}           & 33B                              \\ \midrule
Mistral         & Mistral-7B-Instruct-v0.2\footref{foot:model-mi-s}                & 7B                            \\
                & Mixtral-8x7B-Instruct-v0.1\footref{foot:model-mi-l}                & 47B                         \\  \bottomrule
\end{tabular}
\end{table}

\defnote{foot:model-cl-s}{~\url{https://huggingface.co/meta-llama/Meta-Llama-3-8B-Instruct}}
\defnote{foot:model-cl-l}{~\url{https://huggingface.co/meta-llama/Meta-Llama-3-70B-Instruct}}
\defnote{foot:model-ds-s}{~\url{https://huggingface.co/deepseek-ai/deepseek-coder-6.7b-instruct}}
\defnote{foot:model-ds-l}{~\url{https://huggingface.co/deepseek-ai/deepseek-coder-33b-instruct}}
\defnote{foot:model-mi-s}{~\url{https://huggingface.co/mistralai/Mistral-7B-Instruct-v0.2}}
\defnote{foot:model-mi-l}{~\url{https://huggingface.co/mistralai/Mixtral-8x7B-Instruct-v0.1}}

\subsection{Benchmarks}
We consider two APR benchmarks for evaluating the impact of quantization on LLMs: HumanEval-Java~\cite{jiang2023:impact}, and Defects4J~\cite{just2014:defects4j}.
Both of these benchmarks consist of a set of problems composed of a buggy function, an example fix, and a test suite that can be used to evaluate generated patches. 
\textbf{HumanEval-Java} (HumanEval in the following) is based on the HumanEval benchmark, a collection of 164 coding problems.
While the original benchmark consisted of Python problems, they have been translated to Java, and bugs were injected.
\textbf{Defects4J} offers a collection of Java bugs from open-source projects. 
In particular, we use Defects4J v2.0 and, following practices from previous work~\cite{xiang2024:how, vallecillos-ruiz2025:wisdom}, select the subset of 525 \emph{single-function bugs}, i.e., bugs that exist completely within one single function, from the total set of 835 bugs covered by Defect4J. 

\subsection{Evaluation Metrics}
\label{sec:metrics}
\paragraph{Effectiveness metrics (RQ1)}
To assess the plausibility of a generated patch, we extract the method from the LLM's output and run the related tests.
These tests determine if the code is plausible or not.
We report pass@10 similar to previous works~\cite{afrin2025:quantization,jiang2023:impact,silva2025:repairllama, ruiz2025:art} where a problem is considered to have a plausible solution if any of the 10 generated patches pass all associated tests.
We aggregate this metric per benchmark for each model and configuration.

To capture solved-set consistency between a baseline model and a quantized variant, let $S_b$ and $S_q$ be their solved-problem sets.
We compute the Jaccard Index of the two sets as:
\[
\textit{J}(S_b, S_q) = \frac{|S_b \cap S_q|}{|S_b \cup S_q|}.
\]

However, the Jaccard Index does not clearly indicate the difference between pure count loss due to decreased capabilities and a shift in which problems are solved.
To mitigate this issue, we compute the Jaccard Consistency Rate as:
\[
\textit{JCR} = \frac{\textit{J}(S_b, S_q)}{\min(|S_b|,|S_q|)/\max(|S_b|,|S_q|)}.
\]
The denominator is the maximum Jaccard score achievable given the observed count mismatch.
High JCR indicates that the Jaccard loss is attributable to the count mismatch, therefore indicating that the quantized version solves a subset of the base model solve-problem set.
Low JCR points to a quantized model that solves a different set of problems beyond what the potential raw count reduction would predict.

\paragraph{Efficiency metrics (RQ2)}
We measure inference time (seconds)~\cite{nyamsuren2025:evaluating,daloisio2024:compression,wei2023:greener}, GPU energy consumption (Joules)~\cite{alizadeh2025:language}, in-memory model size (MB)~\cite{kusama2025:how,giagnorio2025:quantizing,daloisio2024:compression,wei2023:greener}, and peak inference memory (MB)~\cite{wei2023:greener}.
In-memory model size is the amount of memory required to load the model parameters before inference, while peak inference memory is the maximum memory usage observed at any point during inference.
For efficiency metrics, lower values are better. 
Section \ref{sec:impl_details} details how these metrics are computed in practice.

\subsection{Statistical \& Trade-offs Analysis}\label{sec:stats}
When reporting the inference time, energy consumption, and peak-inference memory for \textbf{RQ2}, to ensure rigor, we follow performance engineering best practices~\cite{Traini2024, Jangali2023a, Zhang2023a, Laaber2020a}, specifically the approach proposed by Kalibera and Jones~\cite{Kalibera2013a}, to build confidence intervals for the relative change in measurement statistics. In particular, we construct the confidence interval for the median relative change of the metrics computed for each batch using bootstrapping with 10,000 iterations, involving random resampling with replacement~\cite{Kalibera2012a}. The main advantage of this technique, compared to others such as the Wilcoxon test~\cite{woolson2005wilcoxon}, is that it provides a clear and rigorous account of the effectiveness metrics change and the associated uncertainty~\cite{Kalibera2012a,Kalibera2013a,Traini2021a}. For example, this method allows us to state that a quantized model 
requires less memory
than the original model by -30\%$\pm$2\% with 95\% confidence. We consider a difference to be statistically significant if the confidence interval is not greater than the percentage change.\footnote{~Note that, when reporting the results in Section \ref{sec:rq2}, for space reasons, we omit the uncertainty and report only the percentage change, while we mark non-statistically significant variations with an asterisk (*). Nevertheless, full results are available in our replication package.} %

To answer \textbf{RQ3}, we analyze the trade-off between \textit{effectiveness} and \textit{efficiency} in terms of Pareto-dominance~\cite{harman2008search}. Given two solutions $X$ and $Y$ in a multi-objective space, $X$ is said to be \textit{Pareto dominated} by $Y$ if $Y$ is better than $X$ in at least one objective and no worse in the others~\cite{harman2008search}. The set of solutions that are not Pareto dominated by other solutions in the objective space constitutes the Pareto front~\cite{harman2008search}.

\subsection{Implementation Details}\label{sec:impl_details}
All experiments were conducted on an HPC cluster equipped with Nvidia GH200 Hopper GPUs (SM 9.0) and CUDA 12.8.
Jobs were run within Apptainer container images relying on the Transformers inference stack.
For all quantization methods, we installed the recommended libraries from their repositories and applied default configuration parameters.
This design choice aims to reflect the standard, out-of-the-box usage patterns of typical software engineering practitioners.
For each target problem, the model generates 10 candidate patches using a batch size of 1 with stochastic decoding (temperature = 0.1) and 16K maximum context length with fixed seeds to mitigate non-determinism.
During inference, energy consumption is measured via NVML power sampling at 10-millisecond intervals.
Peak inference memory is tracked by adding the maximum delta allocated CUDA memory during the forward pass to the in-memory model load.
Per-problem raw measurements for all metrics are logged and available in the replication package.\footref{footnote:repl-package}

\section{Empirical Results}\label{sec:results}

\subsection{RQ1: Effectiveness}
\subsubsection{RQ1.1: Plausibility}

In RQ1, we evaluate 
the outputs generated by the LLMs and their quantized variants. 
Specifically, we first consider their plausibility (i.e., ability of patches to pass all tests).
Table~\ref{figure:rq1-plausiblityHE} and Table~\ref{figure:rq1-plausiblityDJ} summarize these results for HumanEval and Defects4J respectively, with the best performing method per model highlighted in bold. 

For HumanEval, we observe that the base model never achieves the best plausibility, and is outperformed by at least one quantized model variant for each of the six LLMs. 
The performance difference can be small, with hqq$_4$ (KV) increasing the number of plausible patches from 100 to 101 for Llama-70B, up to an improvement of 19\% for DeepSeek-6.7B and quanto$_4$ (M) (i.e., an increase from 90 to 107 plausible patches).
When considering the quantization methods, we find that hqq (KV) is the best method for Llama, while model quantization methods achieve the best results for Mistral and DeepSeek. 
For both these models, the same quantization configuration is best for the two model sizes (i.e., awq$_4$ for Mistral, quanto$_4$ for DeepSeek). Interestingly, both of these quantize to four-bit integers. 

Unlike HumanEval, the base model achieves the best performance in 1 of 6 cases for Defects4j (Llama-8B), while it is outperformed by quantized models in the remaining five cases.
For Llama-70B, the best performing method remains hqq$_4$(KV).
Similarly to HumanEval, large improvements can be achieved on DeepSeek-6.7B, increasing the number of plausible patches from 43 to 82, an increase of 91\%.
Among the best performing methods are quanto$_8$ for Mistral-7B, and hqq$_8$ (KV) for DeepSeek-33B.

For both benchmarks, we observed that low-bit quantization (2-bit or 3-bit) can lead to poor performance.
For KV Cache quantization, quanto$_4$ always outperforms quanto$_2$, the same holding for hqq$_4$ and hqq$_2$. 
Moreover, aqlm$_2$ and hqq$_3$ model quantization tends to be among the worst performing variants. 
When increasing to 4-bit integers, the results change. In 12 out of 36 cases, we observed that 4-bit quantization is able to perform better or equal to its 8-bit counterpart in terms of the number of plausible patches. 
Additionally, the model family seems to impact the effectiveness of quantization methods. 
The approach hqq(KV) tends to work well for Llama, while awq and quanto(M) achieve good performance for Mistral and DeepSeek, respectively. 

While these results are promising, we observed four cases in which quantization had a significant negative impact on the LLMs' ability to generate compilable code (producing nonsensical outputs or partial answers), leading to 0 or 1 plausible patches over the entire benchmarks.
\begin{table}
\centering
\caption{RQ1: Plausibility on HumanEval. The best performing configuration per model is highlighted in bold.}
\label{figure:rq1-plausiblityHE}
\begin{tabular}{ll|rrrrrrrr}
\toprule
 &  & \multicolumn{2}{c}{\textbf{Llama}} & \multicolumn{2}{c}{\textbf{Mistral}} & \multicolumn{2}{c}{\textbf{DeepSeek}} \\
 &  & \textbf{8B} & \textbf{70B} & \textbf{7B} & \textbf{8x7B} & \textbf{6.7B} & \textbf{33B} \\ \midrule \midrule
 & Base & 74 & 100 & 52 & 88 & 90 & 104 \\ \midrule
\multirow{8}{*}{\rotatebox[origin=c]{90}{Model Weight Quant}} & aqlm$_2$ & 49 & 80 & 4 & 66 & 68 & 33 \\
 & awq$_4$ & 67 & 98 & \textbf{55} & \textbf{93} & 95 & 101 \\
 & bnb$_4$ & 76 & 93 & 51 & 90 & 91 & 104 \\
 & hqq$_3$ & 26 & 1 & 43 & 64 & 56 & 99 \\
 & hqq$_4$ & 66 & 98 & 49 & 82 & 72 & 98 \\
 & hqq$_8$ & \textbf{77} & 98 & 50 & 86 & 92 & 106 \\
 & quanto$_4$ & 68 & 0 & 48 & 79 & \textbf{107} & \textbf{107} \\
 & quanto$_8$ & 72 & 29 & 50 & 85 & 95 & 102 \\ \midrule
\multirow{5}{*}{\rotatebox[origin=c]{90}{KV Quant}} & hqq$_2$ & 43 & 52 & 36 & 52 & 72 & 102 \\
 & hqq$_4$ & 71 & \textbf{101} & 53 & 84 & 92 & 104 \\
 & hqq$_8$ & \textbf{77} & 100 & 48 & 83 & 94 & 101 \\
 & quanto$_2$ & 35 & 23 & 39 & 26 & 62 & 101 \\
 & quanto$_4$ & 71 & 100 & 49 & 87 & 96 & 102 \\ \bottomrule
\end{tabular}
\end{table}

\begin{table}
\centering
\caption{RQ1: Plausibility on Defects4j. The best performing configuration per model is highlighted in bold.}
\label{figure:rq1-plausiblityDJ}
\begin{tabular}{ll|rrrrrrrr}
\toprule
 &  & \multicolumn{2}{c}{\textbf{Llama}} & \multicolumn{2}{c}{\textbf{Mistral}} & \multicolumn{2}{c}{\textbf{DeepSeek}} \\
 &  & \textbf{8B} & \textbf{70B} & \textbf{7B} & \textbf{8x7B} & \textbf{6.7B} & \textbf{33B} \\ \midrule\midrule
 & Base & \textbf{61} & 80 & 35 & 60 & 43 & 81 \\ \midrule
\multirow{8}{*}{\rotatebox[origin=c]{90}{Model Weight Quant}} & aqlm$_2$ & 34 & 52 & 0 & 41 & 19 & 12 \\
 & awq$_4$ & 58 & 81 & 30 & \textbf{76} & 39 & 75 \\
 & bnb$_4$ & 49 & 70 & 39 & 64 & 39 & 83 \\
 & hqq$_3$ & 24 & 3 & 17 & 28 & 66 & 73 \\
 & hqq$_4$ & 48 & 71 & 34 & 39 & 34 & 62 \\
 & hqq$_8$ & 59 & 81 & 34 & 56 & 38 & 86 \\
 & quanto$_4$ & 58 & 0 & 30 & 54 & \textbf{82} & 64 \\
 & quanto$_8$ & 60 & 8 & \textbf{40} & 43 & 44 & 88 \\ \midrule
\multirow{5}{*}{\rotatebox[origin=c]{90}{KV Quant}} & hqq$_2$ & 9 & 21 & 8 & 17 & 19 & 54 \\
 & hqq$_4$ & 60 & \textbf{82} & 32 & 57 & 39 & 83 \\
 & hqq$_8$ & 60 & 80 & 34 & 59 & 40 & \textbf{90} \\
 & quanto$_2$ & 11 & 8 & 9 & 11 & 20 & 56 \\
 & quanto$_4$ & 58 & 71 & 34 & 52 & 40 & 88 \\ \bottomrule
\end{tabular}
\end{table}

\begin{rqanswer}
    \textbf{Answer to RQ1.1:} Contrary to expectations, a quantization of weights to lower precision does not imply a loss in performance. Specifically, there exist at least one quantization method that is able to generate more plausible patches than the base model in 11 out of 12 cases. 
\end{rqanswer}

\begin{figure}
    \centering
    \includegraphics[width=\linewidth]{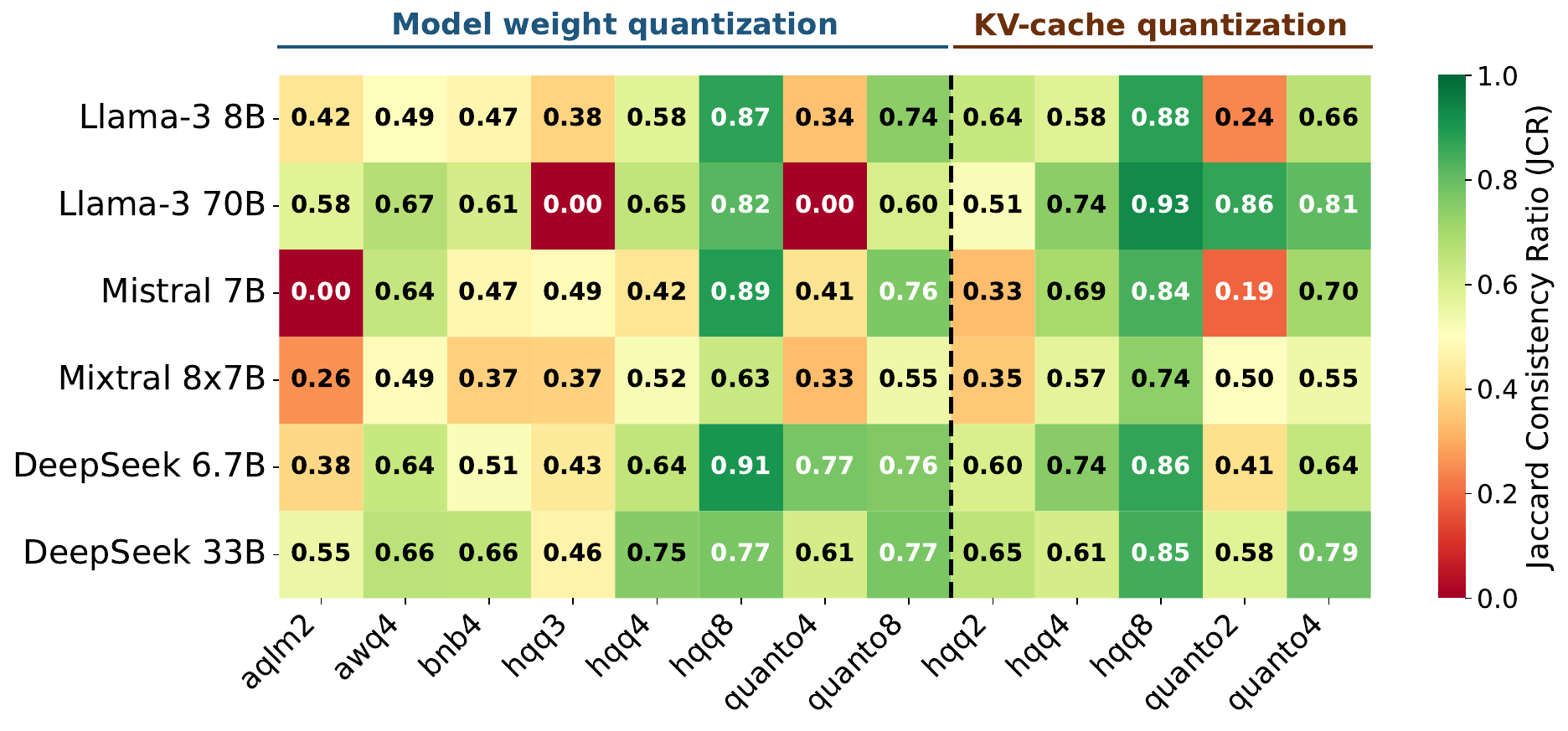}
    \caption{Defects4J JCR heatmap by model and configuration; higher values indicate greater solved-set consistency with the baseline. 
    }
    \label{fig:fidelity-heatmap-jcr-d4j}
\end{figure}

\subsubsection{RQ1.2: Consistency}
In addition to plausibility as a measure of effectiveness, we consider the consistency of quantized models with the base models.
This helps us to understand whether they are able to provide plausible solutions for the same set of problems.
For example, Llama 8B under quanto$_4$(M) solves only 3 problems fewer in Defects4J compared to the baseline (58 vs. 61), but only 29 of those overlap with the baseline.

To analyze this effect, we first calculate the solved-set Jaccard index.
We then use the Jaccard Consistency Ratio (JCR) to interpret the overlap after accounting for raw losses since
JCR distinguishes the overlap loss caused by raw count drops from solved-set drift.
The results for Defects4J are shown in Fig \ref{fig:fidelity-heatmap-jcr-d4j}.
Overall, consistency is higher on HumanEval than on Defects4J, so here we focus the main visualization on Defects4J where the differences are more pronounced.\footnote{~Results for HumanEval are available in our online repository.}
High values identify configurations where the results remain consistent with the baseline, while low values indicate drift beyond what would be expected from a simple count drop as stated in Section \ref{sec:metrics}.

The results indicate that JCR varies widely across configurations with higher-bit configurations generally being more stable.
Even in cases when the number of problems is reduced dramatically, consistency with the baseline results is not guaranteed.
Nevertheless, 8-bit quantization can still introduce non-trivial drift, suggesting that these approaches can alter the capabilities beyond simple ``count loss''.

These results should be taken into account carefully alongside the plausibility count.
Cases with high JCR such as Llama 70B under quanto$_2$ (KV) have a low Jaccard Index due to the count-driven loss (80 $\rightarrow$ 8).
However, JCR helps us to distinguish hqq$_4$ (M) and hqq$_8$ (KV) when applied to Llama. 
The two methods achieve high plausibility counts with hqq$_8$ showing a higher consistency (i.e., 0.88 vs 0.58 for Llama 8B, both variants providing 60 plausible patches).
Consistency also depends on the model family and size.
For example, configurations such as hqq$_2$ have lower values in the Mistral model family, while quanto$_2$ lowered JCR significantly more in smaller models.
This distinction matters since quantized models are often used as substitutes for the full-precision baselines.
In many real-world settings, a loss of overall capability is acceptable only if the behavioural shift is conservative.
Low consistency uncovers cases where the shift is hidden by similar aggregated performance.

\begin{rqanswer}
    \textbf{Answer to RQ1.2:} Similar plausibility counts do not guarantee consistent behaviour after quantization. While higher-bit configurations often remain close to the baseline, extreme quantization can substantially reshuffle the set of bugs repaired. Consistency must be evaluated alongside plausibility to determine whether a quantized model is behaviourally consistent to its original version.
\end{rqanswer}

\subsection{RQ2: Efficiency}\label{sec:rq2}

Tables \ref{tab:inf_time}, \ref{tab:energy}, and \ref{tab:memory_size} report the results observed for inference time, GPU energy consumption and memory footprint, respectively. In each table, the first row reports the median results observed in the base (i.e., non-quantized) model, while the remaining rows show the percentage variations in the median scores by each quantization configuration, computed using the Kalibera and Jones approach (see Section \ref{sec:stats}).

In all tables, for each model and task combination, the worst variations of a quantization configuration compared to the base model are highlighted in \colorbox{black!35}{\textbf{bold}}, while the best variations are \colorbox{gray!25}{\underline{underlined}}.
In the following, we discuss each efficiency metric in detail.

\subsubsection{RQ2.1: Inference Time}\label{sec:inf_time}

\begin{table}[tb]
    \centering
    \caption{RQ2: Inference Time}
    \label{tab:inf_time}
    \resizebox{\linewidth}{!}{
    \begin{tabular}{ll|cccccc}
\toprule
 &  & \multicolumn{2}{c}{\textbf{Llama}} & \multicolumn{2}{c}{\textbf{Mistral}} & \multicolumn{2}{c}{\textbf{DeepSeek}} \\
 &  & \textbf{8B} & \textbf{70B} & \textbf{7B} & \textbf{8x7B} & \textbf{6.7B} & \textbf{33B} \\
\midrule
\midrule
\multicolumn{8}{c}{\textbf{Defects4J}}\\
\midrule
\midrule
& Base (s) & 14.55 & 47.29 & 19.15 & 66.64 & 17.68 & 33.51 \\\midrule
\multirow{8}{*}{\rotatebox[origin=c]{90}{Model Weight Quant}} & aqlm$_2$ & +72.76\% & +204.15\% & +82.82\% & +145.53\% & +138.44\% & +266.38\% \\
& awq$_4$ & +52.32\% & +92.32\% & +68.21\% & +330.92\% & +68.67\% & +44.37\% \\
& bnb$_4$ & +139.57\% & +139.42\% & +145.20\% & +354.42\% & +157.31\% & +129.43\% \\
& hqq$_3$ & \cellcolor{black!25}\textbf{+352.95\%} & \cellcolor{black!25}\textbf{+781.80\%} & \cellcolor{black!25}\textbf{+436.41\%} & \cellcolor{black!25}\textbf{+831.42\%} & \cellcolor{black!25}\textbf{+754.58\%} & \cellcolor{black!25}\textbf{+885.99\%} \\
& hqq$_4$ & +197.74\% & +683.60\% & +207.14\% & +455.15\% & +225.81\% & +436.31\% \\
& hqq$_8$ & +187.36\% & +624.61\% & +174.11\% & +298.35\% & +161.61\% & +550.96\% \\
& quanto$_4$ & +30.89\% & +173.42\% & \cellcolor{gray!25}\underline{+10.05\%} & +102.09\% & +79.25\% & \cellcolor{gray!25}\underline{+24.95\%} \\
& quanto$_8$ & +48.54\% & +187.55\% & +44.13\% & +188.08\% & +55.25\% & +156.71\% \\\midrule
\multirow{5}{*}{\rotatebox[origin=c]{90}{KV Quant}} & hqq$_2$ & +82.87\% & +149.87\% & +112.81\% & +290.07\% & +162.27\% & +103.00\% \\
& hqq$_4$ & +48.91\% & \cellcolor{gray!25}\underline{+63.46\%} & +54.88\% & +115.02\% & +98.69\% & +62.99\% \\
& hqq$_8$ & \cellcolor{gray!25}\underline{+27.54\%} & +97.84\% & +29.99\% & \cellcolor{gray!25}\underline{+81.85\%} & \cellcolor{gray!25}\underline{+58.72\%} & +33.23\% \\
& quanto$_2$ & +52.27\% & +134.87\% & +90.30\% & +234.26\% & +80.22\% & +63.71\% \\
& quanto$_4$ & +36.21\% & +89.56\% & +30.99\% & +184.49\% & +46.86\% & +39.81\% \\
\midrule
\midrule
\multicolumn{8}{c}{\textbf{HumanEval}}\\
\midrule
\midrule
 & Base (s) & 6.24 & 22.64 & 9.24 & 34.69 & 8.20 & 17.19 \\
\midrule
\multirow{8}{*}{\rotatebox[origin=c]{90}{Model Weight Quant}} & aqlm$_2$ & +81.62\% & +182.01\% & +72.93\% & +67.27\% & +107.04\% & +286.43\% \\
& awq$_4$ & +45.56\% & +61.75\% & +62.25\% & +124.13\% & +83.97\% & +59.71\% \\
& bnb$_4$ & +129.71\% & +148.03\% & +126.87\% & +206.73\% & +148.86\% & +128.94\% \\
& hqq$_3$ & \cellcolor{black!25}\textbf{+427.97\%} & \cellcolor{black!25}\textbf{+714.31\%} & \cellcolor{black!25}\textbf{+406.33\%} & \cellcolor{black!25}\textbf{+493.50\%} & \cellcolor{black!25}\textbf{+560.33\%} & \cellcolor{black!25}\textbf{+807.70\%} \\
& hqq$_4$ & +193.25\% & +668.03\% & +195.17\% & +232.95\% & +205.98\% & +308.43\% \\
& hqq$_8$ & +188.08\% & +587.09\% & +164.70\% & +183.86\% & +158.17\% & +513.00\% \\
& quanto$_4$ & \cellcolor{gray!25}\underline{+11.66\%} & +157.17\% & \cellcolor{gray!25}\underline{+10.95\%} & \cellcolor{gray!25}\underline{+18.65\%} & +40.03\% & \cellcolor{gray!25}\underline{+11.88\%} \\
& quanto$_8$ & +44.44\% & +190.39\% & +42.46\% & +95.58\% & +49.76\% & +154.28\% \\\midrule
\multirow{5}{*}{\rotatebox[origin=c]{90}{KV Quant}} & hqq$_2$ & +60.04\% & +95.56\% & +79.03\% & +148.18\% & +111.45\% & +94.81\% \\
& hqq$_4$ & +48.92\% & +50.68\% & +40.04\% & +44.80\% & +64.02\% & +44.35\% \\
& hqq$_8$ & +24.69\% & \cellcolor{gray!25}\underline{+46.82\%} & +27.44\% & +43.90\% & \cellcolor{gray!25}\underline{+25.66\%} & +21.96\% \\
& quanto$_2$ & +30.74\% & +120.07\% & +60.56\% & +84.78\% & +57.24\% & +71.27\% \\
& quanto$_4$ & +33.79\% & +56.88\% & +29.62\% & +31.81\% & +20.37\% & +29.68\% \\
\bottomrule
\end{tabular}}
\end{table}

Table \ref{tab:inf_time} reports the results observed for inference time (measured in seconds) under Defects4J and HumanEval.

First, we note how all quantization configurations increase the inference time of the base model, despite its architecture and the complexity of the underlying task. This result is in line with what has been observed previously on CodeBERT~\cite{daloisio2024:compression}. A possible explanation is that GPUs and processing libraries are heavily optimized for full-precision floating-point operations (e.g., 32-bit); hence, quantized models, with their default hyperparameters, may not fully take advantage of these optimizations and experience an inference slowdown~\cite{ganesh2021compressing} (see also Section \ref{sec:threats}). 
Another pattern that is observed in all models and in both tasks is that hqq$_3$ emerges as the quantization configuration that increases the inference time the most, with variations up to +885.99\% observed in DeepSeek 33B on Defect4J. This could be explained by the nature of hqq quantization, which has a higher impact on inference time when the batch size is small (as in our case, where the batch size is one)~\cite{zhang2025efficient}. Moreover, we observe that, in general, hqq quantization has a much worse impact on inference time when applied to the whole model weights rather than only the KV cache. Conversely, quanto %
has a different impact depending on the task and the model. 

In addition, different patterns are observed in quantization configurations that affect inference time less. First, quanto$_4$(M) is the configuration that increases inference time the least in 6 out of 12 (50\%) cases, followed by hqq$_8$(KV), which is best in 41.7\% of the cases. Another interesting pattern is observed in hqq(M), where increasing the precision from 3 to 4 bits significantly reduces the increase in inference time in all models except Llama 70B. %

\begin{rqanswer}
    \textbf{Answer to RQ2.1:} All quantization configurations increase inference time, with hqq$_3$ being the worst configuration in all models and tasks, with increases up to +885.99\%. On the contrary, quanto$_4$(M) is the configuration that increases inference time the least in 50\% scenarios.
\end{rqanswer}

\subsubsection{RQ2.2: Energy Consumption}\label{sec:energy}

\begin{table}[tb!]
    \centering    
    \caption{RQ2: Energy Consumption}
    \label{tab:energy}
    \resizebox{\linewidth}{!}{\begin{tabular}{ll|cccccc}
        \toprule
         &  & \multicolumn{2}{c}{\textbf{Llama}} & \multicolumn{2}{c}{\textbf{Mistral}} & \multicolumn{2}{c}{\textbf{DeepSeek}} \\
         &  & \textbf{8B} & \textbf{70B} & \textbf{7B} & \textbf{8x7B} & \textbf{6.7B} & \textbf{33B} \\
        \midrule
        \midrule
        \multicolumn{8}{c}{\textbf{Defect4J}}\\
        \midrule
        \midrule
        & Base (J) & 3509.35 & 23420.40 & 4822.97 & 22925.64 & 6549.48 & 10687.82 \\\midrule
        \multirow{8}{*}{\rotatebox[origin=c]{90}{Model Weight Quant}} & aqlm$_2$ & +129.33\% & +173.97\% & +100.47\% & +103.99\% & +165.53\% & +402.02\% \\
       & awq$_4$ & +47.37\% & +69.12\% & +54.27\% & +320.95\% & \cellcolor{gray!25}\underline{+45.95\%} & +39.98\% \\
        & bnb$_4$ & +138.97\% & +153.03\% & +164.45\% & +352.23\% & +126.29\% & +162.39\% \\
        & hqq$_3$ & \cellcolor{black!25}\textbf{+725.85\%} & \cellcolor{black!25}\textbf{+1112.00\%} & \cellcolor{black!25}\textbf{+814.93\%} & \cellcolor{black!25}\textbf{+1306.08\%} & \cellcolor{black!25}\textbf{+1098.52\%} & \cellcolor{black!25}\textbf{+1511.28\%} \\
        & hqq$_4$ & +551.68\% & +988.22\% & +590.16\% & +798.15\% & +490.07\% & +809.64\% \\
        & hqq$_8$ & +541.71\% & +919.76\% & +523.45\% & +568.66\% & +359.22\% & +1019.11\% \\
        & quanto$_4$ & \cellcolor{gray!25}\underline{+33.54\%} & +158.15\% & \cellcolor{gray!25}\underline{+14.07\%} & +95.50\% & +79.39\% & +49.16\% \\
        & quanto$_8$ & +149.12\% & +259.67\% & +137.04\% & +249.09\% & +109.17\% & +283.76\% \\\midrule
        \multirow{5}{*}{\rotatebox[origin=c]{90}{KV Quant}} & hqq$_2$ & +91.51\% & +106.61\% & +127.11\% & +286.80\% & +216.21\% & +92.08\% \\
        & hqq$_4$ & +58.10\% & +44.80\% & +72.39\% & +114.73\% & +147.91\% & +62.87\% \\
        & hqq$_8$ & +35.31\% & \cellcolor{gray!25}\underline{+63.63\%} & +41.27\% & \cellcolor{gray!25}\underline{+80.03\%} & +93.87\% & +44.69\% \\
        & quanto$_2$ & +55.89\% & +88.79\% & +85.80\% & +227.56\% & +113.03\% & +53.33\% \\
        & quanto$_4$ & +40.26\% & +65.53\% & +36.45\% & +187.24\% & +72.73\% & \cellcolor{gray!25}\underline{+29.88\%} \\
         \midrule
        \midrule
        \multicolumn{8}{c}{\textbf{HumanEval}}\\
        \midrule
        \midrule
         & Base (J) & 1334.20 & 9651.93 & 1938.63 & 11160.20 & 1949.78 & 4817.71 \\\midrule
         \multirow{8}{*}{\rotatebox[origin=c]{90}{Model Weight Quant}} & aqlm$_2$ & +137.50\% & +180.13\% & +86.92\% & +47.36\% & +151.94\% & +443.87\% \\
        & awq$_4$ & +34.13\% & +67.79\% & +59.22\% & +142.25\% & +87.05\% & +52.89\% \\
        & bnb$_4$ & +131.55\% & +180.53\% & +129.55\% & +214.91\% & +150.20\% & +173.95\% \\
        & hqq$_3$ & \cellcolor{black!25}\textbf{+887.23\%} & \cellcolor{black!25}\textbf{+1121.82\%} & \cellcolor{black!25}\textbf{+825.89\%} & \cellcolor{black!25}\textbf{+857.83\%} & \cellcolor{black!25}\textbf{+962.46\%} & \cellcolor{black!25}\textbf{+1524.46\%} \\
        & hqq$_4$ & +579.98\% & +1102.18\% & +577.11\% & +502.70\% & +555.36\% & +659.27\% \\
        & hqq$_8$ & +567.26\% & +987.44\% & +508.52\% & +408.56\% & +448.09\% & +1035.15\% \\
        & quanto$_4$ & \cellcolor{gray!25}\underline{+18.38\%} & +167.06\% & \cellcolor{gray!25}\underline{+14.34\%} & \cellcolor{gray!25}\underline{+20.26\%} & +51.90\% & +42.10\% \\
        & quanto$_8$ & +152.54\% & +309.46\% & +141.83\% & +163.50\% & +135.79\% & +309.66\% \\\midrule
         \multirow{5}{*}{\rotatebox[origin=c]{90}{KV Quant}} & hqq$_2$ & +34.61\% & +87.85\% & +62.97\% & +143.18\% & +153.40\% & +73.47\% \\
        & hqq$_4$ & +43.19\% & \cellcolor{gray!25}\underline{+35.85\%} & +41.73\% & +41.84\% & +106.31\% & +35.61\% \\
        & hqq$_8$ & +27.01\% & +38.52\% & +23.73\% & +38.51\% & +54.55\% & \cellcolor{gray!25}\underline{+21.67\%} \\
        & quanto$_2$ & +23.29\% & +104.08\% & +46.75\% & +80.06\% & +73.74\% & +59.72\% \\
        & quanto$_4$ & +23.61\% & +44.32\% & +27.79\% & +38.89\% & \cellcolor{gray!25}\underline{+39.10\%} & +27.66\% \\
        \bottomrule
        \end{tabular}}
\end{table}

Table \ref{tab:energy} reports the results observed for GPU energy consumption (measured in Joules) under Defects4J and HumanEval.

First, it is worth noting that energy consumption and inference time are highly positively correlated, as reported by a statistically significant Spearman correlation $>0.95$ for all models analysed~\cite{spearman1961proof}.\footnote{~We employed the non-parametric Spearman correlation instead of the parametric Pearson correlation since the Shapiro-Wilk test rejected the null hypothesis of normally distributed data for all efficiency metrics.} Therefore, the patterns observed in Table \ref{tab:energy} are similar to those discussed in Section \ref{sec:inf_time} for inference time. Specifically, all quantization configurations increase the energy consumption of the base model. Among the different configurations, hqq$_3$(M) is the one mostly impacting energy consumption in both tasks, with increments up to +1524.46\% for DeepSeek 33B under HumanEval. 

Following the high correlation between inference time and energy, quanto$_4$ (applied either on model weights or KV cache) and hqq$_4$(KV) and hqq$_8$(KV) are the quantization configurations that impact energy consumption the least. 
However, although the high correlation with inference time, we observe a single exception on awq$_4$, which emerges at the configuration with the least impact on energy consumption in one case. %

\begin{rqanswer}
    \textbf{Answer to RQ2.2:} Inference time and energy consumption are highly positively correlated and exhibit a similar trend. hqq$_3$(M) is the configuration that mostly increases energy consumption with increases up to +1524.46\%. quanto$_4$(M) is instead the model that increases energy consumption the least in 5 out of 12 (41.7\%) scenarios.
\end{rqanswer}

\subsubsection{RQ2.3: Memory Footprint}

\begin{table}[tb]
    \centering
    \caption{RQ2: Memory Footprint}
    \label{tab:memory_size}
    \begin{subtable}{\linewidth}
    \centering
    \resizebox{\linewidth}{!}{\begin{tabular}{l|cccccc}
         \toprule
          & \multicolumn{2}{c}{\textbf{Llama}} & \multicolumn{2}{c}{\textbf{Mistral}} & \multicolumn{2}{c}{\textbf{DeepSeek}} \\
          & \textbf{8B} & \textbf{70B} & \textbf{7B} & \textbf{8x7B} & \textbf{6.7B} & \textbf{33B} \\
            \midrule
            \midrule
            Base (MB) & 15316.51 & 134570.52 & 13812.51 & 89078.51 & 12856.51 & 63596.71\\\midrule
            aqlm$_2$ & \cellcolor{gray!25}\underline{-74.57\%} & \cellcolor{gray!25}\underline{-84.47\%} & \cellcolor{gray!25}\underline{-84.30\%} & \cellcolor{gray!25}\underline{-85.98\%} & \cellcolor{gray!25}\underline{-82.30\%} & \cellcolor{gray!25}\underline{-85.59\%} \\
            awq$_4$ & -64.34\% & -71.81\% & -71.34\% & -73.61\% & -70.90\% & -72.93\% \\
            bnb$_4$ & -64.48\% & -71.97\% & -71.50\% & -73.78\% & -71.29\% & -73.14\% \\
            hqq$_3$ & -68.58\% & -72.49\% & -72.11\% & -73.47\% & -71.81\% & -73.13\% \\
            hqq$_4$ & -63.95\% & -67.72\% & -67.25\% & -68.51\% & -67.04\% & -68.09\% \\
            hqq$_8$ & \cellcolor{black!25}\textbf{-40.66\%} & \cellcolor{black!25}\textbf{-43.10\%} & \cellcolor{black!25}\textbf{-42.73\%} & \cellcolor{black!25}\textbf{-43.63\%} & \cellcolor{black!25}\textbf{-42.06\%} & \cellcolor{black!25}\textbf{-43.05\%} \\
            quanto$_4$ & -63.67\% & -71.17\% & -70.60\% & -72.96\% & -70.43\% & -72.32\% \\
            quanto$_8$ & -43.44\% & -48.50\% & -48.17\% & -49.70\% & -47.91\% & -49.28\% \\
            \bottomrule\\
    \end{tabular}}
    \caption{In-Memory Model Size}
    \label{tab:model_size}
\end{subtable}

\begin{subtable}{\linewidth}
    \centering
        \resizebox{\linewidth}{!}{\begin{tabular}{ll|cccccc}
         \toprule
         &  & \multicolumn{2}{c}{\textbf{Llama}} & \multicolumn{2}{c}{\textbf{Mistral}} & \multicolumn{2}{c}{\textbf{DeepSeek}} \\
         &  & \textbf{8B} & \textbf{70B} & \textbf{7B} & \textbf{8x7B} & \textbf{6.7B} & \textbf{33B} \\
            \midrule
            \midrule
        \multicolumn{8}{c}{\textbf{Defects4J}}\\
        \midrule
        \midrule
         & Base (MB) & 16541.66 & 138086.81 & 15318.08 & 90661.61 & 18219.35 & 66389.10 \\\midrule
        \multirow{8}{*}{\rotatebox[origin=c]{90}{Model Weight Quant}} & aqlm$_2$ & \cellcolor{gray!25}\underline{-69.16\%} & \cellcolor{gray!25}\underline{-82.55\%} & \cellcolor{gray!25}\underline{-76.11\%} & \cellcolor{gray!25}\underline{-84.53\%} & \cellcolor{gray!25}\underline{-60.48\%} & \cellcolor{gray!25}\underline{-81.40\%} \\
        &awq$_4$ & -60.00\% & -69.70\% & -64.76\% & -72.13\% & -52.88\% & -70.09\% \\
        &bnb$_4$ & -59.67\% & -69.84\% & -64.46\% & -72.31\% & -52.36\% & -69.89\% \\
        &hqq$_3$ & -28.07\% & -60.62\% & -55.68\% & -69.90\% & -40.61\% & -66.10\% \\
        &hqq$_4$ & -23.65\% & -55.89\% & -51.60\% & -65.06\% & -41.02\% & -61.69\% \\
        & hqq$_8$ & -1.76\% & -31.87\% & -29.29\% & -40.61\% & -23.08\% & -37.35\% \\
        &quanto$_4$ & -59.06\% & -69.27\% & -64.32\% & -71.67\% & -49.11\% & -69.54\% \\
        &quanto$_8$ & -40.17\% & -46.94\% & -43.33\% & -48.65\% & -35.48\% & -46.96\% \\\midrule
        \multirow{5}{*}{\rotatebox[origin=c]{90}{KV Quant}} & hqq$_2$ & -3.13\% & -0.87\% & -3.96\% & -0.39\% & -12.47\% & -2.21\% \\
        & hqq$_4$ & -2.72\% & -0.74\% & -3.47\% & -0.34\% & -10.21\% & -1.93\% \\
        & hqq$_8$ & \cellcolor{black!25}\textbf{-1.54\%} & \cellcolor{black!25}\textbf{-0.48\%} & \cellcolor{black!25}\textbf{-1.82\%} & \cellcolor{black!25}\textbf{-0.04\%} & \cellcolor{black!25}\textbf{-5.21\%} & \cellcolor{black!25}\textbf{-1.23\%} \\
        & quanto$_2$ & -3.34\% & -0.87\% & -4.29\% & -0.64\% & -16.09\% & -2.27\% \\
        & quanto$_4$ & -2.92\% & -0.74\% & -3.74\% & -0.58\% & -13.92\% & -2.00\% \\
        \midrule
        \midrule
        \multicolumn{8}{c}{\textbf{HumanEval}}\\
        \midrule
        \midrule
         & Base (MB) & 15880.70 & 136079.52 & 14480.08 & 89770.04 & 15337.96 & 64868.68 \\\midrule
        \multirow{8}{*}{\rotatebox[origin=c]{90}{Model Weight Quant}} & aqlm$_2$ & \cellcolor{gray!25}\underline{-71.64\%} & \cellcolor{gray!25}\underline{-83.43\%} & \cellcolor{gray!25}\underline{-79.84\%} & \cellcolor{gray!25}\underline{-85.24\%} & \cellcolor{gray!25}\underline{-68.46\%} & \cellcolor{gray!25}\underline{-83.17\%} \\
        &awq$_4$ & -62.19\% & -70.53\% & -67.97\% & -72.79\% & -42.85\% & -69.25\% \\
        &bnb$_4$ & -61.90\% & -70.66\% & -67.51\% & -72.97\% & -59.20\% & -71.34\% \\
        &hqq$_3$ & -28.62\% & -61.14\% & -58.45\% & -70.51\% & -48.77\% & -67.65\% \\
        &hqq$_4$ & -24.35\% & -56.36\% & -53.93\% & -65.60\% & -46.42\% & -63.10\% \\
        &hqq$_8$ &-1.75\% & -32.01\% & -30.52\% & -40.91\% & -25.80\% & -38.16\% \\
        &quanto$_4$ & -61.43\% & -70.15\% & -67.44\% & -72.38\% & -57.68\% & -71.03\% \\
        &quanto$_8$ & -41.56\% & -47.44\% & -45.28\% & -49.08\% & -39.85\% & -47.93\% \\
        \midrule
        \multirow{5}{*}{\rotatebox[origin=c]{90}{KV Quant}} & hqq$_2$ & -1.47\% & -0.39\% & -1.79\% & -0.08\% & -6.35\% & -0.97\% \\
        & hqq$_4$ & -1.17\% & -0.34\% & -1.52\% & -0.09\% & -5.71\% & -0.86\% \\
        & hqq$_8$ & \cellcolor{black!25}\textbf{-0.73\%} & \cellcolor{black!25}\textbf{-0.22\%} & \cellcolor{black!25}\textbf{-0.84\%} & \cellcolor{black!25}\textbf{0.00\%\textsuperscript{*}} & \cellcolor{black!25}\textbf{-3.24\%} & \cellcolor{black!25}\textbf{-0.53\%} \\
        & quanto$_2$ & -1.70\% & -0.40\% & -2.29\% & -0.31\% & -8.69\% & -1.10\% \\
        & quanto$_4$ & -1.45\% & -0.34\% & -1.99\% & -0.26\% & -7.85\% & -0.96\% \\
        \bottomrule\\
    \end{tabular}}
    \caption{Memory footprint\footref{foot:stats}}
    \label{tab:inference_mem}
\end{subtable}
\vspace*{-3ex}
\end{table}

Results concerning memory footprint are reported in Table \ref{tab:memory_size}.
In particular, Table \ref{tab:model_size} shows the variations observed concerning the in-memory model size, while Table \ref{tab:inference_mem} reports the variations observed on the memory footprint. It is worth noting that KV cache quantization methods change the precision of the KV cache at inference time and do not touch the actual model weights. Therefore, they do not alter the model's actual in-memory size, but only the memory required for inference. For this reason and due to space constraints, KV cache quantization methods are omitted from Table \ref{tab:model_size}. At the same time, we do not distinguish between Defects4J and HumanEval when reporting results in Table \ref{tab:model_size}, since the in-memory size of a model is independent of the underlying task.

Table \ref{tab:model_size} shows that all model weight quantization methods are effective in reducing the in-memory size of the base models. 
A clear trend emerges where lower weight precision yields greater memory reduction. Accordingly, aqlm$_2$ (\texttt{int2}) achieves the greatest memory savings across all models, while hqq$_8$ (\texttt{int8}) yields the minimum.
These results are expected and align with previous research~\cite{daloisio2024:compression,giagnorio2025:quantizing}.

The pattern reflects on the memory footprint reported in Table \ref{tab:inference_mem}. In fact, aqlm$_2$ remains the most effective quantization configuration for all models and tasks, indicating that the reduction in in-memory model size does not lead to an increase in the memory required for inference. However, we observe an overall low memory reduction achieved by KV cache quantization methods compared to model weight ones. Indeed, hqq$_8$(KV) quantization is the configuration that least reduces the memory footprint of all models in both tasks. 
This behaviour could be explained by the fact that the size of the KV cache depends on the length of the input prompt~\cite{hooper2024kvquant}. 
Therefore, if the input prompt length is not high, the KV cache will be small, and its reduction will be less effective~\cite{hooper2024kvquant}. 
In fact, from Table \ref{tab:inference_mem}, we observe that the memory reduction achieved by KV cache quantization is slightly higher in Defects4J, which has overall longer input prompts. 
Still, all KV cache quantization methods are dominated by model-weight ones.

\defnote{foot:stats}{~Non-statistically significant results are marked with an asterisk (*)}

Finally, it is worth noting that, unlike inference time and energy consumption, the memory footprint is not highly correlated with any of the other efficiency metrics, with Spearman correlation values $<0.2$ across all models analysed.

\begin{rqanswer}
    \textbf{Answer to RQ2.3:} All quantization methods reduce the memory footprint of the base models, with aqlm$_2$ and hqq$_8$ being the best and worst configuration in models and tasks, respectively. 
    In our study,
    model-weight quantization methods outperform all KV cache quantization methods in memory footprint reduction.
\end{rqanswer}

\subsection{RQ3: Effectiveness-Efficiency Trade-offs}

\begin{table}
\centering
\scriptsize
\caption{RQ3: Percentage of Pareto-dominated quantization configurations. %
}
\begin{tabular}{l|cccccc}
\toprule
          & \multicolumn{2}{c}{\textbf{Llama}} & \multicolumn{2}{c}{\textbf{Mistral}} & \multicolumn{2}{c}{\textbf{DeepSeek}} \\
          & \textbf{8B} & \textbf{70B} & \textbf{7B} & \textbf{8x7B} & \textbf{6.7B} & \textbf{33B} \\
\midrule
\midrule
\textbf{Defects4J} & 46.15\% & 61.54\% & 38.46\% & 46.15\% & 46.15\% & 46.15\% \\
\textbf{HumanEval} & 38.46\% & 53.85\% & 38.46\% & 53.85\% & 61.54\% & 46.15\% \\
\bottomrule
\end{tabular}
\label{tab:dominated}
\end{table}

\begin{figure}[b]
    \centering
    \includegraphics[width=\linewidth]{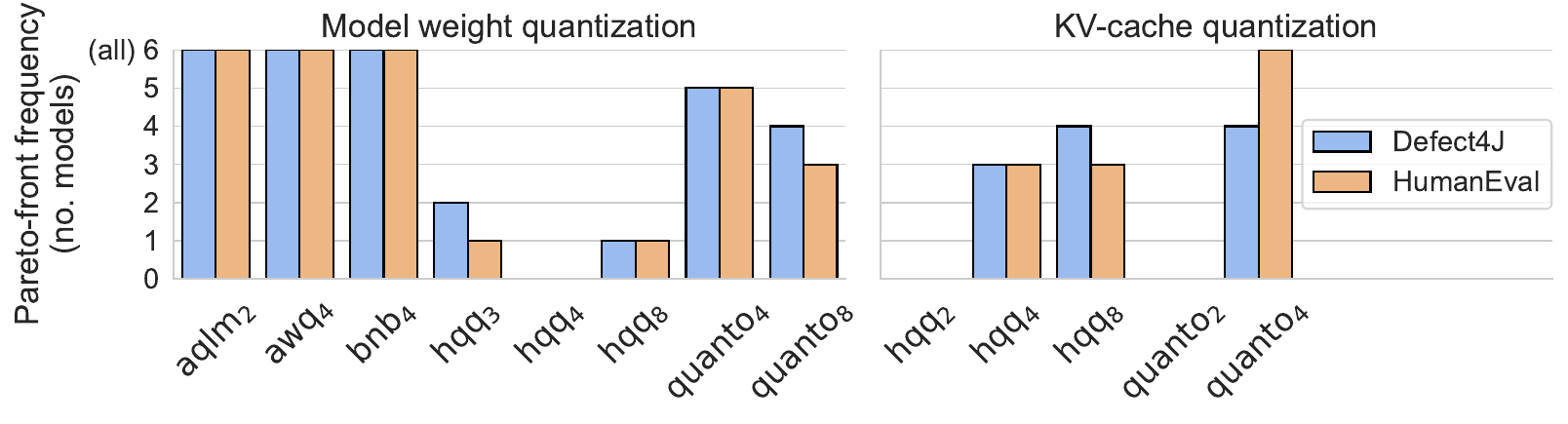}
    \caption{RQ3: Number of models in which each quantization configuration appears on the Pareto front.}
    \label{fig:paretofront_freq}
\end{figure}

\begin{figure*}
    \centering
    \includegraphics[width=1.03\linewidth,trim=7pt 0 0 0,clip]{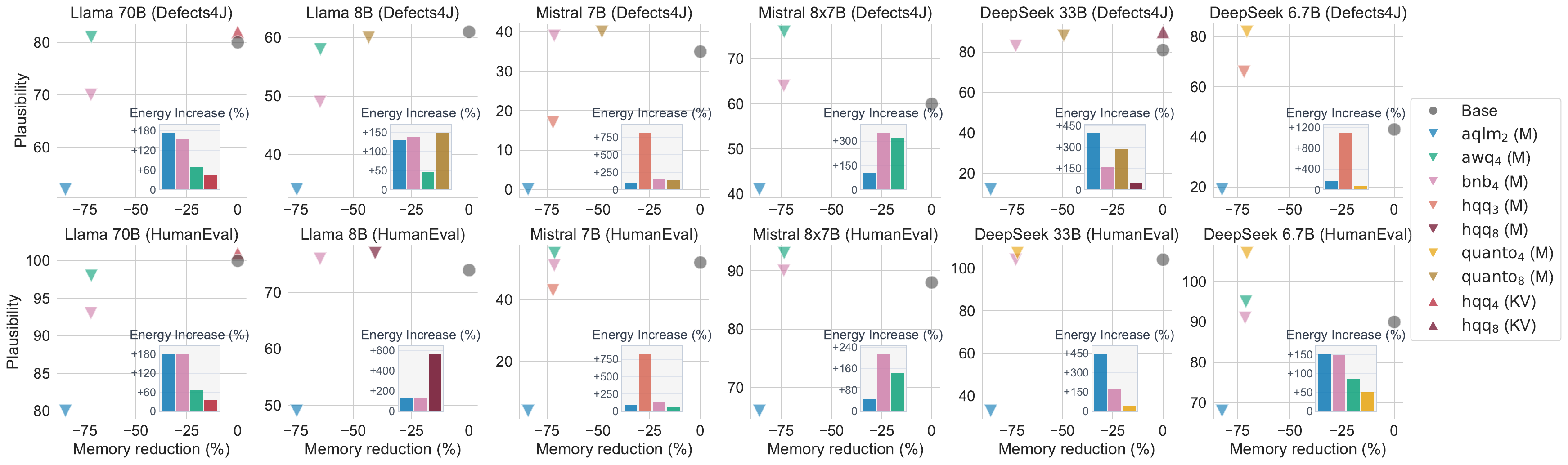}
    \caption{RQ3: Pareto frontier (+ base model) for each model and benchmark, considering plausibility (pass@10) and in-memory model size. The x-axis shows in-memory model size reduction (lower is better), while the y-axis shows the plausibility (higher is better). The bar plots depict the energy increase of the quantization configurations relative to the base model. (M) indicates model weight quantization, while (KV) indicates KV-cache quantization.}
    \label{fig:pareto_scatter}
\end{figure*}

Here, we assess the effectiveness-efficiency trade-offs of different quantization configurations and compare them with one another.
Table \ref{tab:dominated} shows the percentages of Pareto-dominated quantization configurations for each model and benchmark, considering all efficiency metrics (inference time, energy consumption, in-memory model size, and peak inference memory) as well as the effectiveness metric (plausibility). A quantization configuration is considered Pareto-dominated by a different quantization configuration or the base model if the latter is at least as good on every metric and strictly better on at least one.
On average across all model–benchmark pairs, 48\% of quantization configurations are Pareto-dominated, suggesting a non-trivial share of suboptimal configurations for which model efficiency (resp., effectiveness) can be improved without sacrificing effectiveness (resp., efficiency) simply by switching to a different quantization configuration.
In the worst cases -- Llama 70B on Defect4J and DeepSeek 6.7B on HumanEval -- 61.54\% of configurations (8 out of 13) are Pareto-dominated.
Even in the best cases -- Llama 8B on HumanEval and Mistral 7B on both benchmarks -- 38.46\% of configurations (5 out of 13) are Pareto-dominated.
Overall, these results underscore the need to select quantization configurations carefully to avoid suboptimal trade-offs.

Figure \ref{fig:paretofront_freq} shows how frequently different quantization configurations appear on the Pareto front across models, considering all efficiency metrics as well as plausibility.
We find that some quantization configurations, such as aqlm$_2$, awq$_4$, and bnb$_4$, appear on the Pareto front for all six models, suggesting they are unlikely to be dominated by other configurations regardless of the language model used. For KV-cache quantization, the most frequent configuration is quanto$_4$, with Pareto-front frequencies of 6 on HumanEval and 4 on Defect4J.
In contrast, other quantization configurations are more frequently Pareto-dominated. For example, hqq$_3$ appears on the Pareto front for only two models on Defect4J and one model on HumanEval, while hqq$_8$ appears for only one model on each benchmark. The worst quantization configurations in this regard are hqq$_4$ (M), hqq$_2$ (KV), and quanto$_2$ (KV), which are systematically dominated by other configurations across all models and benchmarks.
The remaining configurations 
(quanto$_4$(M), quanto$_8$(M), hqq$_4$(KV) and hqq$_8$(KV)) 
show mixed results, with Pareto-front frequencies ranging from 3 to 5.

Figure \ref{fig:pareto_scatter} shows the configurations on the Pareto frontier in terms of in-memory model size and APR effectiveness (plausibility), along with the base model for each model–benchmark pair. The x-axis shows the relative memory reduction due to quantization, while the y-axis shows the plausibility score. 
Observe that distinct configurations can have nearly similar memory reductions, yet have very different plausibility scores.
We consider only model-size memory rather than peak inference memory, because model size better reflects the average memory footprint under typical usage and has been used in prior studies to assess the trade-off between effectiveness and memory footprint~\cite{giagnorio2025:quantizing,daloisio2024:compression}. The overlaid bar plot shows the percentage energy increase due to quantization for each configuration on the frontier. In this analysis, we omit inference time since it is highly correlated with energy consumption (see Section \ref{sec:impl_details}).
As expected, model quantization with aqlm$_2$ achieves the highest memory reduction, though with a significant reduction in APR effectiveness, consistently showing the lowest plausibility score.
Model quantization based on awq$_4$ and bnb$_4$ shows the most balanced memory-effectiveness trade-offs across models and benchmarks. In that, we observe awq$_4$ tends to lead to higher APR effectiveness and lower energy increase compared to bnb$_4$, e.g., Llama 70B (both benchmarks), Llama 8B (HumanEval), Mistral 7B (HumanEval), Mistral 8x7B (both benchmarks), and DeepSeek 6.7B (HumanEval).
Interestingly, some configurations appear to work better only on specific benchmarks or models. For instance, on Defect4J, quanto$_8$ achieves comparable or better plausibility scores while substantially reducing model size (-50\%) in 3 out of 5 models. Similarly, quanto$_4$ yields the best plausibility for DeepSeek 6.7B (both benchmarks) and DeepSeek 33B (HumanEval), while reducing memory usage by about 70\% and providing the lowest increase in energy consumption among the Pareto-optimal configurations.

\begin{rqanswer}
    \textbf{Answer to RQ3:} Overall, 48\% of the evaluated quantization configurations are Pareto-dominated by alternative settings. While some configurations (e.g., awq$_4$) generally provide a more favorable balance between APR effectiveness and efficiency performance, others appear sensitive to particular models or benchmarks (e.g., quanto$_4$ and quanto$_8$) in achieving competitive trade-offs.
\end{rqanswer}

\section{Discussion}\label{sec:discussion}

This section discusses some implications of this study. %
A key insight for practitioners seeking to apply quantization in APR is that quantization does not necessarily degrade effectiveness. When the quantization configuration is chosen appropriately for the target model and the program-repair context, quantized models can achieve APR effectiveness comparable to that of the base model. However, even when aggregate repair effectiveness remains similar, quantization can induce shifts in repair behavior. For instance, some bugs that the base model can fix may no longer be fixed after quantization, while the quantized model may successfully repair other bugs that the base model cannot. Practitioners should therefore anticipate potential capability drift when deploying APR systems based on quantized models.

Another relevant consideration is that efficiency is not necessarily improved by quantization. In fact, in our experiments, both inference time and energy consumption consistently increased for quantized models. These results correspond to previous findings on CodeBERT~\cite{daloisio2024:compression}.
This counterintuitive observation may be explained by the dequantization overhead introduced by quantization (i.e., frequent type conversions) and by the fact that the integer kernels used by quantized models are less optimized on GPUs than the fp16/bf16 kernels used by unquantized models~\cite{ganesh2021compressing}. 
Consequently, if practitioners face strict constraints on inference time or energy consumption, our findings indicate that they should avoid using quantized models for APR. If the constraints are less strict, selecting an appropriate quantization configuration for the specific model and program-repair context may lead to smaller efficiency penalties, although some increase in inference time and energy consumption should still be expected (e.g., +10\% inference time and +14\% energy consumption in the best cases).

If the practitioner's primary objective is to reduce memory footprint, quantization can yield substantial reductions in in-memory model size (ranging from -42\% to -86\%), in some cases with marginal or no impact on APR effectiveness, provided that the configuration is chosen appropriately. However, our experimental results indicate that an uninformed, arbitrary choice of quantization configuration can easily lead to suboptimal outcomes, since nearly half of the evaluated configurations are Pareto-dominated by alternative configurations in terms of effectiveness and efficiency. Therefore, selecting an appropriate quantization configuration is critical to fully exploit these benefits. Some configurations appear to be safer default choices for practitioners (e.g., awq$_4$(M)), as they better balance memory reductions with limited losses in APR effectiveness
across our evaluations.
That said, those seeking to fully exploit memory-efficiency gains with limited (or no) degradation in APR effectiveness should tune the choice of quantization configuration to the specific model and repair context.

\section{Threats to Validity}\label{sec:threats}
A threat to the \textbf{internal validity} of our study is related to the selection of hyperparameters for applying quantization methods and generating outputs with the LLMs. 
These settings may not be optimal for every configuration.
To address this, we used the default parameter for quantization methods to represent out-of-the-box usage and identical settings for LLM inference.
These settings may not be optimal for every configuration and could change the performance of the methods.
Furthermore, stochastic behaviour is also present as an internal threat.
We did not repeat the full experiment with multiple independent runs since the current evaluation already required approximately 900 GPU-hours on GH200 GPUs.
We expect the conclusions, such as memory reduction, inference time, energy, and solved-set drift, to remain.
However, the exact number of problems solved and Pareto fronts may vary across runs.
Another threat to internal validity is data leakage, which occurs when data from the evaluation benchmark has been part of the model training. 
To alleviate this, we consider two benchmarks, with HumanEval-Java~\cite{jiang2023:impact} being more recently introduced.
We focus our analysis on relative changes between base and quantized models which remain a valid signal even if absolute baselines are inflated,
however, it does not fully remove leakage concerns.
We also note that, although we cannot eliminate the risk of data leakage, the problems remain challenging, as evidenced by the fact that our best models solve only 17\% of the Defects4J problems.
Another internal threat concerns the measurement of non-functional metrics (e.g., inference time, energy consumption, and peak inference memory usage), which may be affected by other background processes. Although we cannot fully ensure the absence of other background processes, we mitigated this threat by relying on SLURM to run each job in an isolated environment and by employing established techniques from performance engineering to mitigate potential outliers~\cite{daloisio2024:compression,Traini2024}.

\textbf{External validity} addresses the generalizability of our results. 
This can be related to other models, quantization methods, programming language, and hardware settings. 
We mitigate this threat by including multiple quantization methods (with varying levels of precision) as well as six LLMs of different sizes, 
and while both benchmarks are for Java, they include a wide range of bug types.
Still, effectiveness results such as plausibility and solved-set drift may differ for other programming languages.
On the other hand, efficiency results are less tied to Java, but they still depend on prompt length, model architecture, and inference setup.
Furthermore,
our non-functional measurements were obtained on NVIDIA GH200 Hopper GPUs.
Observed latency and energy penalties may not easily extrapolate to resource-constrained consumer GPUs.
Unoptimized low-bit quantization configurations can bottleneck the inference, increasing decoding time and, consequently, total energy.
An experimentation with further hardware settings would be a valuable avenue for future work. 

A threat to \textbf{construct validity} is the use of plausibility (i.e., passing a test suite) as a proxy for correctness for generated patches. 
While this is a practical choice, the quality of results depends on the comprehensiveness of the test suites, and in some cases, patches overfit to the test suite.
One potential way to combat this is the manual inspection of patches.
However, this process is prone to subjectivity~\cite{wang2020:automated} and recent works on APR showed that only a small fraction of plausible end up being not correct~\cite{kang2024:explainable, ruiz2025:art}. 
Regarding non-functional metrics, our measurements are bounded by the inference phase.
We focus on GPU-side consumption under a batch size of 1, omitting system-level costs.
Inference memory is also dependent on batch size; therefore, our findings may differ under high-throughput serving environments.

\section{Conclusion and Future Work}\label{sec:conclusion}

In this paper, we studied how quantization affects the effectiveness and efficiency of LLMs for APR. We conducted an empirical evaluation on six LLMs and 13 quantization configurations, derived from five methods, targeting either model weights or the KV cache.
Our results show that quantization introduces model- and context-dependent trade-offs for APR.
When the quantization configuration is selected appropriately, quantized models can often match (or even exceed) the repair effectiveness of their full-precision counterparts, while providing notable memory footprint gains.
However, quantization can also be detrimental to efficiency dimensions, particularly energy consumption and inference time.

Our findings motivate future work on (i) 
expanding the study to other types of quantization and combinations of them, (ii) understanding the shift induced by quantization on the problems solved, and (iii)
routing policies that dispatch buggy programs to the most suitable quantized (or full-precision) model to jointly optimize effectiveness and efficiency in multi-LLM repair pipelines. We plan to pursue these two directions as the core of our future research agenda.

\section*{Acknowledgments}

\noindent
This work is partially supported by \SoBigDataIPAck. This article is also partially funded by Università degli Studi dell'Aquila under the Call for Proposals for Fundamental research and Early-career research grants - Year 2026.
In addition, it is partially supported by the Research Council of Norway through the secureIT project (IKTPLUSS \#288787), and by the European Union through the Horizon Europe Marie Sk\l{}odowska-Curie Actions (\#101151798).
The empirical evaluation made use of the Experimental Infrastructure for Exploration of Exascale Computing (eX3), 
financially supported by the Research Council of Norway under contract \#270053, as well as on resources provided by
Sigma2 - the National Infrastructure for High-Performance Computing and Data Storage in Norway.

\balance
\printbibliography

\end{document}